\documentclass[journal,twoside,web]{ieeecolor}
\usepackage{generic}
\usepackage{cite}
\usepackage{amsmath,amssymb,amsfonts}
\usepackage{algorithmic}
\usepackage{graphicx}
\usepackage{algorithm,algorithmic}
\usepackage{hyperref}
\hypersetup{hidelinks=true}
\usepackage{textcomp}
\usepackage{url}
\usepackage{xcolor}
\usepackage{float}
\usepackage{hyperref}
\usepackage{stfloats}
\usepackage{amssymb}
\usepackage{latexsym}
\usepackage{amsmath}
\usepackage{makecell}
\usepackage{booktabs}
\usepackage{arydshln}
\usepackage{array}
\usepackage{multirow}
\usepackage{upgreek}
\usepackage{threeparttable}

\def\BibTeX{{\rm B\kern-.05em{\sc i\kern-.025em b}\kern-.08em
    T\kern-.1667em\lower.7ex\hbox{E}\kern-.125emX}}
\begin{document}
\title{Spiral Scanning and Self-Supervised Image Reconstruction Enable Ultra-Sparse Sampling Multispectral Photoacoustic Tomography}
\author{Yutian Zhong, Xiaoming Zhang, Zongxin Mo, Shuangyang Zhang, Wufan Chen, \IEEEmembership{Senior Member, IEEE}, and Li Qi 
\thanks{This work was supported in part by National Natural Science Foundation of China (62371220), Guangdong Basic and Applied Basic Research Foundation (2021A1515012542, 2022A1515011748), and Guangdong Pearl River Talented Young Scholar Program
(2017GC010282). (\textit{Corresponding authors: Li Qi.})}
\thanks{The authors are with School of Biomedical Engineering, Southern Medical University, Guangzhou, Guangdong, China; Guangdong Provincial Key Laboratory of Medical Image Processing, Southern Medical University, Guangzhou, Guangdong, China; and Guangdong Province Engineering Laboratory for Medical Imaging and Diagnostic Technology, Southern Medical University, Guangzhou, Guangdong, China. (e-mail: ytzhong.smu@qq.com; 1182740969@qq.com; hhapk1@163.com; syzhang@smu.edu.cn; chenwf@smu.edu.cn; qili@smu.edu.cn).}}

\maketitle

\begin{abstract}
Multispectral photoacoustic tomography (PAT) is an imaging modality that utilizes the photoacoustic effect to achieve non-invasive and high-contrast imaging of internal tissues. However, the hardware cost and computational demand of a multispectral PAT system consisting of up to thousands of detectors are huge. To address this challenge, we propose an ultra-sparse spiral sampling strategy for multispectral PAT, which we named U3S-PAT. Our strategy employs a sparse ring-shaped transducer that, when switching excitation wavelengths, simultaneously rotates and translates. This creates a spiral scanning pattern with multispectral angle-interlaced sampling. To solve the highly ill-conditioned image reconstruction problem, we propose a self-supervised learning method that is able to introduce structural information shared during spiral scanning. We simulate the proposed U3S-PAT method on a commercial PAT system and conduct \textit{in vivo} animal experiments to verify its performance. The results show that even with a sparse sampling rate as low as 1/30, our U3S-PAT strategy achieves similar reconstruction and spectral unmixing accuracy as non-spiral dense sampling. Given its ability to dramatically reduce the time required for three-dimensional multispectral scanning, our U3S-PAT strategy has the potential to perform volumetric molecular imaging of dynamic biological activities.
\end{abstract}

\begin{IEEEkeywords}
Sparsely sampled image reconstruction, photoacoustic tomography, implicit neural representation, prior embedding.
\end{IEEEkeywords}

\section{Introduction}
\label{sec:introduction}

\IEEEPARstart{P}{hotoacoustic} tomography (PAT) is a novel imaging approach that merges the high contrast of optical imaging with the deep imaging depth of ultrasound to produce cross-sectional images within tissue \cite{b1}--\cite{b6}. Multispectral PAT captures images at multiple wavelengths, enabling the identification of concentration and location of photoacoustic absorbers based on spectral unmixing. This capability aids in the precise visualization of various tissue components, such as fat, blood, and tumors \cite{b7}--\cite{b14}.

Conventional PAT imaging system usually employs a linear or ring-shaped transducer for data acquisition, and the acquired images are 2D cross-sectional images. With linear translation scanning, a 3D volume image can be obtained, as shown in Fig.1 (a). To further perform multispectral imaging, one image is obtained per excitation wavelength. The total data amount and acquisition time increase with the number of wavelengths and slices. Therefore, multispectral and 3D PAT imaging has a heavy time cost and is very inefficient.

In practical applications, reducing the data acquisition burden can be accomplished through sparse sampling, which is usually implemented by reducing the total number of transducer elements \cite{b15}--\cite{b17}. It cuts down the cost of the transducer and signal acquisition equipment. However, as the number of detectors decreases, the angular coverage becomes sparser, posing challenges for achieving high-quality image reconstruction. Novel image reconstruction algorithms including iterative model-based approaches \cite{b18}--\cite{b21} and recently introduced deep-learning-based (DL-based) techniques \cite{b22,b23} have been devised to address this issue, yet preserving detailed structural information while avoiding excessive smoothing remains challenging. Currently, sparse sampling PAT systems only achieve a minimum sparse sampling rate of $<$ 1/4, in order to maintain acceptable image reconstruction quality. Notably, these sparse sampling techniques do not reduce the total image acquisition time for multispectral imaging.

\begin{figure*}[!h]
\centering
\includegraphics[scale=.35]{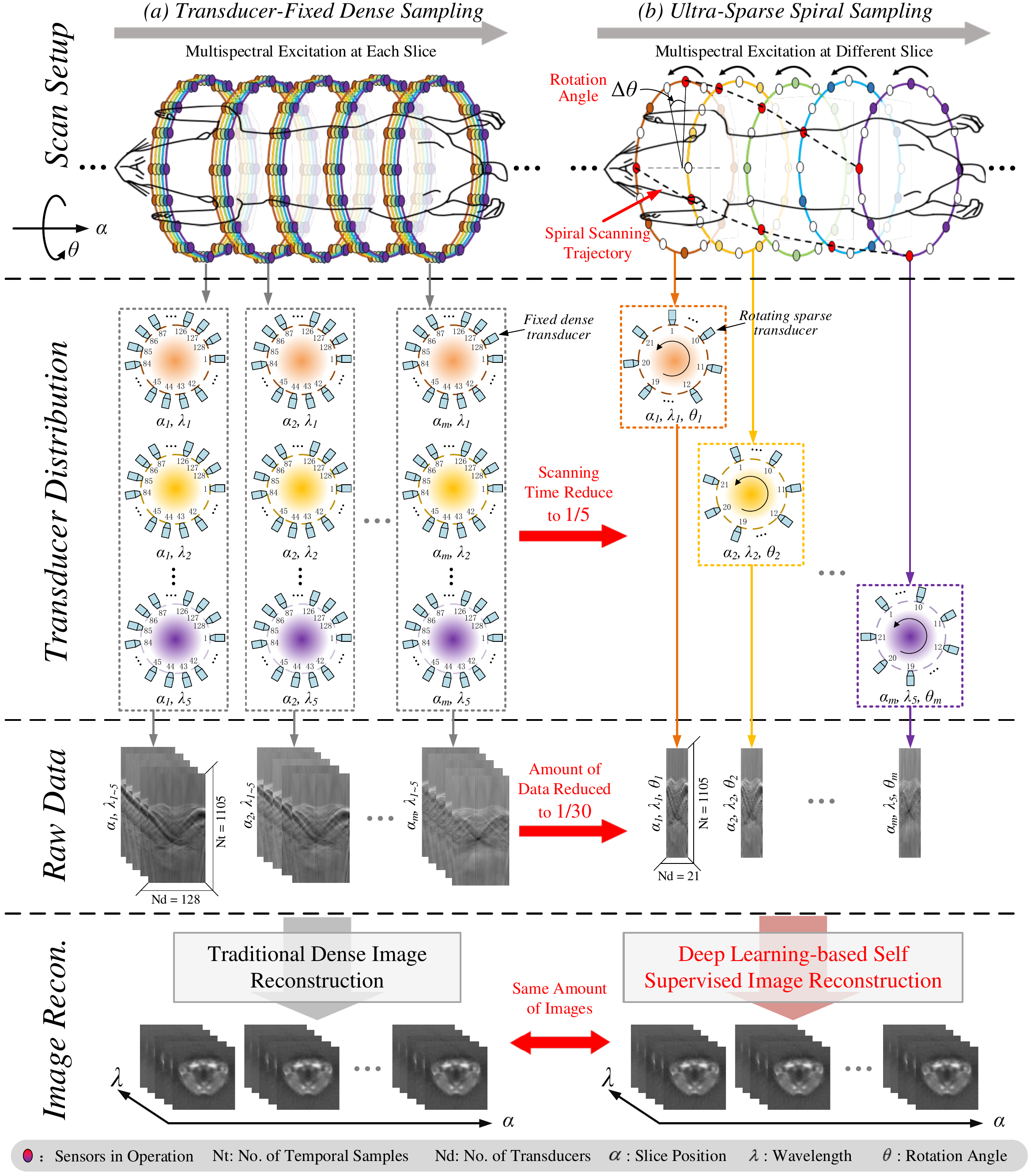}
\caption{(a) Traditional transducer fixed dense sampling (DS) strategy: the detection array consists of a large number of elements (e.g. $\rm Nd = 128$). A dense set of signal is acquired at each slice at multiple wavelengths. (b) The proposed ultra-sparse spiral sampling (U3S-PAT) strategy: the sparse transducer (e.g. $\rm Nd = 21$) rotates and translates linearly while switching excitation wavelength, thus creating a spiral scanning trajectory. The scanning time and amount of data can be dramatically reduced.}
\end{figure*}

This brings us to an important question: is there any space to further reduce the sparse sampling rate? One way to do this is to make use of the redundant spatial-spectral information during scanning. As a proof of concept, our group recently proposed a multispectral sparse sampled PAT method, named interlaced sparse sampling PAT (ISS-PAT) \cite{b24}, where the sparse ring-shaped transducer array rotates by an angle for each wavelength switch. This scheme leverages the shared information during each laser excitation, it offers image reconstruction quality similar to dense sampling at a sparse sampling rate as low as 1/8. Yet, ISS-PAT can only image a single slice, and for 3D imaging accomplished by linear scanning, the scanning time will still be increased accordingly.

Except for utilizing shared spectral information, we can also leverage the redundant spatial information shared during 3D scanning for efficient imaging. For example, spiral scanning, which rotates and translates the detector at the same time, is a highly efficient imaging concept widely adopted in clinical X-ray computed tomography (X-ray CT) imaging \cite{b25}--\cite{b28}. Recently, spiral scanning has been introduced to photoacoustic imaging. Initially proposed in \cite{b29}, spiral volumetric PAT based on a high-end spherical array transducer has been used to characterize brown adipose tissue \cite{b30} and generate panoramic images of mice \cite{b31}. In these studies, spiral scanning is used to compensate for the limited imaging view of the spherical transducer. Their data acquisition is still dense sampling, and the cost of a spherical transducer with a multi-channel synchronized data acquisition system is very high. Therefore, optimizing the scanning time and system burdens is a problem to be solved, especially in multispectral 3D imaging.

For cross-sectional PAT with a ring-shaped transducer, the axial displacement between two slices is on the order of sub-millimeters. Therefore, there is a wealth of information shared between slices that may enable sparser sampling. Inspired by this, we propose to combine the technique of spiral scanning with multispectral interlaced sparse sampling to further reduce the sparse sampling rate. As shown in Fig.1 (b), our approach, which is dubbed ultra-sparse spiral sampling PAT or U3S-PAT, involves the simultaneous rotation and translation of the sparse ring-shaped transducer each time the excitation wavelength is changed. By doing so, our U3S-PAT facilitates multi-wavelength and multi-slice cross-sectional imaging that covers the entire region of interest with much fewer measurements. To solve the highly ill-conditioned image reconstruction problem, we develop a novel self-supervised image reconstruction model that does not require any ground truth for network training and only relies on intrinsic information of the raw PA signal. Provided by our unique U3S-PAT scanning strategy, we incorporate a structural prior image that fuses the redundancy information among slices and wavelengths to guide image reconstruction. To assess the performance of the proposed U3S-PAT strategy, we simulate the acquisition of spiral scanning data on a commercial PAT system and conduct \textit{in vivo} animal experiments. The results demonstrate that even with a sparse sampling rate as low as 1/30, the reconstruction image quality and spectral unmixing accuracy achieved using our U3S-PAT are similar to those obtained with full-angle, all-wavelength dense sampling.

\section{Methods}
\subsection{Multispectral ultra-sparse spiral sampling PAT}
Our U3S-PAT method is based on a cross-sectional multispectral PAT system with a ring-shaped array transducer. Currently, most of popular PAT systems are based on the dense sampling strategy [Fig.1 (a)], where the transducer array consists of a large number of elements (e.g., the total number of elements $\rm Nd = 128$) to ensure dense angular coverage. Either the transducer or the imaging target is linearly translated for imaging of different slices. At each slice position, multiple sets of signal are obtained for multispectral imaging. Therefore, the dense sampling scheme inevitably generates a large amount of data, i.e., the total size of acquired data is $\rm W \times Nd \times Nt \times M$, where $\rm Nt$ is the number of temporal samples obtained by each detector, $\rm W$ is the number of excitation wavelengths in each slice and $\rm M$ is the number of scanned slices.

In our U3S-PAT strategy [Fig.1 (b)], a sparse transducer array is used (e.g. $\rm Nd = 21$). During scanning, each time the excitation wavelength is switched from one another, in the meantime, the transducer rotates for a small angle around the imaged object and translates a specified distance along the axial direction. Therefore, the slice position, scanning wavelength, and angle of the transducer are changed per image simultaneously. A spiral scanning trajectory is formed for the transducer. To make full use of the angular spacing between two adjacent transducer elements, the transducer is rotated such that each measurement is evenly distributed. That means, letting $\varphi$ be the angle between two adjacent elements, the angle $\mathrm{\Delta}\theta$ rotated by the transducer as the laser transits from one wavelength to another is $\mathrm{\Delta}\theta = \varphi/ \rm{W}$. Therefore, the rotation angle at the tomographic position $\alpha_{m}\left( {m \in \left\{ {1,2,\cdots \rm{M}} \right\}} \right)$ is denoted as $\theta_{m}$, where $\theta_{m} = m \times \mathrm{\Delta}\theta$. 

As a comparison, under our U3S-PAT strategy, the total size of the acquired data is only 
$r \rm{\times Nd \times Nt \times M}$, where $r$ is the downsampling rate of transducer elements. Therefore, compared to dense sampling, the total sparse sampling rate is $1/r\rm{W}$. For example, for $\rm W = 5$, compared to dense sampling using 128 elements, we can reduce the amount of data to 1/30 with 21 elements, i.e. $r \approx 1/6$. More importantly, our U3S-PAT strategy reduces multispectral scanning time during 3D imaging. To image the same number of slices, the time it takes for traditional transducer-fixed PAT to perform $\rm W$ wavelength imaging can be reduced to only $\rm 1/W$. Since only one wavelength is used at each slice, the imaging speed of U3S-PAT is limited by the wavelength switching time. Therefore, our U3S-PAT may enable 3D multispectral imaging of dynamic scenes such as the beating heart, thus making possible the visualization of oxyhemoglobin ($\rm {HbO}_{2}$) and deoxyhemoglobin ($\rm {Hb}$) within the heart during cardiac cycles.

\subsection{Image reconstruction based on implicit neural representation}
The U3S-PAT principle results in ultra-sparse data. Therefore, reconstructing high-quality images of all the slices at all wavelengths from such few measurements is a significant challenge. Recent developments in deep learning (DL) techniques offer powerful tools to tackle this kind of problem, but well-developed supervised DL methods require high-quality ground truth images which is difficult to obtain in sparse sampled PAT. Alternatively, neural representations that do not require ground truth data for training have recently garnered significant attention \cite{b32}--\cite{b35}, with application on CT and MR reconstruction recently demonstrated \cite{b32}. Inspired by these works, rather than learning end-to-end mapping, we propose a prior-embedded multi-layer perceptron (MLP) to learn the implicit neural representation of the target image from U3S-PAT data without any densely sampled reference.

\begin{figure*}[!h]
\centering
\includegraphics[scale=.45]{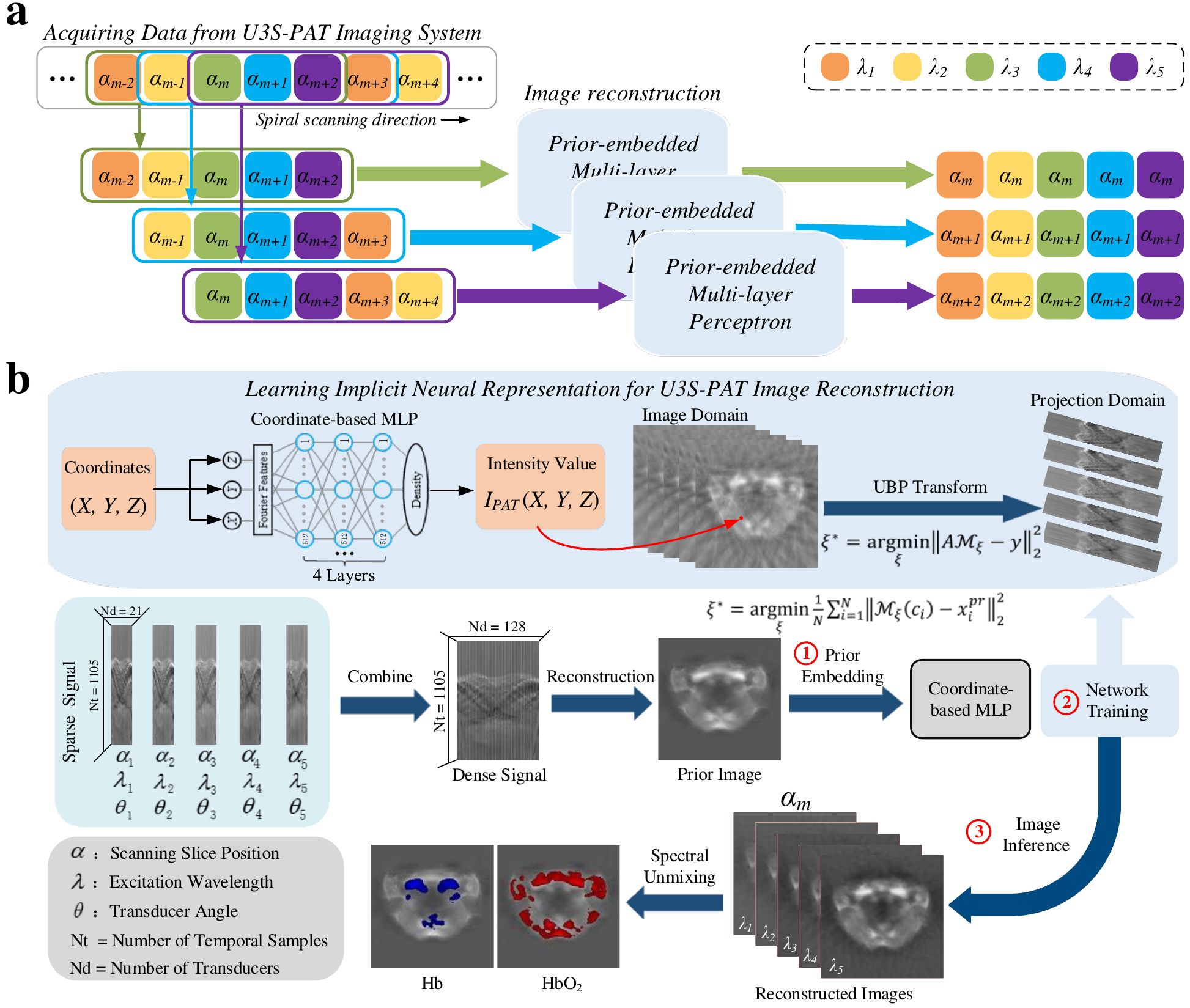}
\caption{(a) Schematic of the image acquisition and reconstruction procedure of U3S-PAT. The prior-embedded MLP network takes the detected signal from five adjacent slices as input, and reconstructs five images at the center slice position corresponding to the five excitation wavelengths. (b) Schematic diagram of the implicit neural representation-based image reconstruction model. The wavelength and angle interlaced raw data is combined to form a dense signal and then reconstructed into a structural prior image. The image is embedded into the MLP network to guide multi-wavelength image reconstruction.}
\end{figure*}

Since the transducer in cross-sectional PAT has a certain slice thickness limited by its focus, there is redundant information shared between adjacent slices if the translation distance is smaller than the slice thickness. Making use of this information should be able to improve image reconstruction. Fig.2 shows the principle of our self-supervised image reconstruction method. As shown in Fig.2 (a), in our U3S-PAT strategy, only one wavelength is used at each slice, and each slice corresponds to a single wavelength. Specifically, since the imaging wavelength loops through 
$\lambda_{1}$ to $\lambda_{\rm{W}}$, the wavelength index at $\alpha_{m}$ is given by $\omega = \left( {m - 1} \right)\rm{mod}(\rm{W}) + 1$, where $\rm{mod}$ is the modulus operator and $\omega = 1,2,\cdots \rm{W}$. To reconstruct images at all wavelengths, the remaining spectral information can be induced from neighboring slices. Consequently, the input to our network model is a set of data of continuous slices, and the total number of slices is determined by the number of excitation wavelengths, i.e. $\rm{W}$, such that all wavelengths are covered. The output of the network is consisting of multi-wavelength images of the center slice position. Thus, the main function of the network model is to transfer the missing spectral information from adjacent slices to the center slice.

Fig.2 (b) shows the schematic of the reconstruction network at a given tomographic position when 
$\rm W = 5$. Considering the interlaced scanning principle of U3S-PAT, we can combine all the acquired spirally scanned signal to form a dense set of signal with a high angular coverage density. From the combined signal we can then reconstruct a fusion image that integrates information from different wavelengths, transducer rotation angles, and tomographic positions. Therefore, this fusion image has rich, high-quality structural information, and thus can be used as prior knowledge to guide image reconstruction. To do so, the weights of the MLP network are embedded with internal information from the prior image to serve as an initialization for the search target image representation. Starting from this pre-embedded initialization, the network is optimized to identify the optimal points in the function space based solely on sparsely sampled measurements.

Our method involves three steps to derive the final images. Firstly, the prior image is encoded as an implicit neural representation by embedding the entire spatial image field into network parameters. Secondly, using the prior-embedded network as initialization, for training network to learn the reconstruction of multi-wavelength data, we utilized the sparse sampling data obtained at the given slice as well as its neighboring slices acquired at different wavelengths to perform the constrained reconstruction. Finally, by traversing all spatial coordinates in the image space, the learned MLP generates the reconstructed images, i.e. images of the center slice at $\lambda_{1}$ to $\lambda_{5}$. By performing the above steps at each tomographic position, a complete set of multi-wavelength images at all slices can be obtained.

\subsubsection{Network architecture}
The neural network in our approach is implemented as a 4-layer MLP network comprising 512 neural nodes in width. In each layer except the final layer, we employed a periodic activation function, known for its effectiveness in capturing fine details in signals \cite{b35}. The image is represented as a continuous function within the neural network. We define the network, denoted as $\mathcal{M}_{\xi}$, with the parameter $\xi$, as follows:

\begin{equation} 
\left. \mathcal{M}_{\xi}:c\rightarrow v~with~c \in \lbrack 0,1)^{n},~v \in \mathbb{R} \right.
\end{equation}

\noindent The network function $\mathcal{M}_{\xi}$ takes the normalized spatial coordinates $c$ as input and produces the corresponding intensity value $v$ as output. By mapping coordinates to image intensities, the network function $\mathcal{M}_{\xi}$ encapsulates the internal information of the entire image within its parameters $\xi$. Thus, we consider the network structure $\mathcal{M}_{\xi}$ with parameters $\xi$ as a neural representation of the images because it captures the essential characteristics of the images.

\subsubsection{Prior embedding}
We incorporate the prior images $x^{pr}$ into the network during initialization. The coordinate-based MLP network $\mathcal{M}_{\xi}$ is employed to map the spatial coordinates to their corresponding intensity values in the prior images $x^{pr}$. Mathematically, this can be expressed as $\left. \mathcal{M}_{\xi}:c_{i}\rightarrow x_{i}^{pr} \right.$, where $i$ denotes the index of coordinates in the image's spatial field. With a total of $N$ pixels in the image denoted as $\left\{ {c_{i},x_{i}^{pr}} \right\}_{i = 1}^{N}$. The initially randomized MLP network $\mathcal{M}_{\xi}$ is optimized using the following objective function:

\begin{equation} 
\xi^{*} = \underset{\xi}{argmin}\frac{1}{N}{\sum_{i = 1}^{N}\left\| {\mathcal{M}_{\xi}\left( c_{i} \right) - x_{i}^{pr}} \right\|_{2}^{2}}
\end{equation}

Following the optimization process, $\mathcal{M}_{\xi^{*}}$ is updated to encode the internal information of the prior images $x^{pr}$. This encoding is achieved by combining the optimized network parameters $\xi^{*}$ with $\mathcal{M}_{\xi^{*}}$. To maintain clarity and distinction, we denote the resulting MLP network as $\mathcal{M}^{pr}$, representing the prior-embedded MLP network. In other words, the prior images $x^{pr}$ are equivalent to the output of this embedded MLP network, denoted as $x^{pr} = \mathcal{M}_{\xi^{*}} = \mathcal{M}^{pr}$.

\subsubsection{Network training}
Utilizing the pre-embedded MLP network $\mathcal{M}^{pr}$ and the given sparse measurements $y$, where $y$ includes sparse sampling data for each tomographic position. We aim to train the network to acquire the neural representation of target images. The target images $x$, being dependent on the coordinate-based MLP network $\mathcal{M}_{\xi}$ and its parameter $\xi$, are utilized to define the data term as $\underset{x}{min}\mathcal{~}\mathcal{E}\left( {Ax,y} \right) = \underset{\xi}{min}\mathcal{~}\mathcal{E}\left( A\mathcal{M}_{\xi},y \right)$, where $\mathcal{E}\left( {Ax,y} \right)$ is a data term used to measure the error between the $Ax$ and $y$ variables to ensure that the data is consistent with the sensor measurements. Matrix $A$ represents the forward model of the imaging system. The function $\mathcal{E}$ is a distance metric such as L1 or L2 norm. This formulation enables the optimization of the MLP parameter space to translate into the optimization of the image space. The network $\mathcal{M}_{\xi}$  undergoes training through the minimization of the L2 parametric loss, leveraging the initialization provided by the prior embedding network $\mathcal{M}^{pr}$. Consequently, the optimization objective for the network $\mathcal{M}_{\xi}$ can be expressed as follows:

\begin{equation} 
\xi^{*} = \underset{\xi}{argmin}\left\| {A\mathcal{M}_{\xi} - y} \right\|_{2}^{2},~x^{*} = \mathcal{M}_{\xi^{*}}
\end{equation}

\noindent The forward model $A$ utilized in this paper is the universal back-projection (UBP) transform, which possesses differentiability properties. By combining this with the differentiable orthorectified model of the PAT imaging system, a link between the image space and the transducer space is established. 

To reconstruct a complete set of multi-wavelength images, we employ a joint constraint reconstruction using the sparse sampling data obtained from not only the given slice position, but also from neighboring slices acquired at multiple wavelengths. Therefore, for reconstructing multi-wavelength images of tomographic position $\alpha_{m}$, since its corresponding excitation wavelength is $\lambda_{\omega}$, where $\omega = \left( {m - 1} \right)mod(\rm {W}) + 1$, the reconstruction loss for the $\lambda_{\omega}$ image can be expressed as:

\begin{equation} 
\xi^{*} = \underset{\xi}{argmin}\left\| {A\mathcal{M}_{\xi} - y_{\alpha_{m}}} \right\|_{2}^{2},~x^{*} = \mathcal{M}_{\xi^{*}}
\end{equation}

For the excitation wavelengths other than $\lambda_{\omega}$, the image reconstruction loss is:

\begin{equation} 
\begin{aligned} 
&\xi^{*} = \underset{\xi}{argmin}\left( {\left\| {A\mathcal{M}_{\xi} - y_{\alpha_{m}}} \right\|_{2}^{2} + \delta\left\| {A\mathcal{M}_{\xi} - y_{\alpha_{m + k}}} \right\|_{2}^{2}} \right), \\
&~x^{*} = \mathcal{M}_{\xi^{*}}
\end{aligned} 
\end{equation}

\noindent Where $k \rm \in \left\{ {1 - \frac{W + 1}{2},2 - \frac{W + 1}{2},\cdots,W - \frac{W + 1}{2}} \right\}$, $k \neq 0$, and $\delta$ is a hyperparameter.

\subsubsection{Images inference}
Once the network has undergone training, the reconstructed images can be obtained by deducing the trained network across all spatial coordinates within the image field. Mathematically, this can be represented as $x^{*}:\left\{ {c_{i},\mathcal{M}_{\xi^{*}}\left( c_{i} \right)} \right\}_{i = 1}^{N}$, where $i$ denotes the index of the coordinates. The resulting intensity values at these coordinates form the reconstructed multi-wavelength images $x^{*}$ at each tomographic position.

\section{Experimental setup}
\subsection{U3S-PAT data acquisition}
All experiments are conducted using a commercial multispectral PAT platform (MSOT inVision128, iThera Medical GmbH, Germany). The system employs a ring transducer array comprising 128 elements and a radius of 40.5 mm. The transducer elements have a central frequency of 5 MHz. The system is equipped with a pulsed optical parametric oscillator (OPO) laser (SpotLight 600, InnoLas, Germany), capable of emitting pulsed laser light ranging from 680 nm to 980 nm. The laser pulses are emitted through 10 irradiation ports, ensuring a uniform distribution of laser energy across the surface of the imaged object. The laser operates at a pulse repetition frequency of 10 Hz, with a wavelength switching time of 30 ms. When irradiating a cylindrical sample with a diameter of 20.0 mm, the laser creates a ring-shaped illumination with a width of 8.0 mm on the sample's surface, as shown in Fig.3 (a).

\begin{figure}[ht]
\centering
\includegraphics[scale=.37]{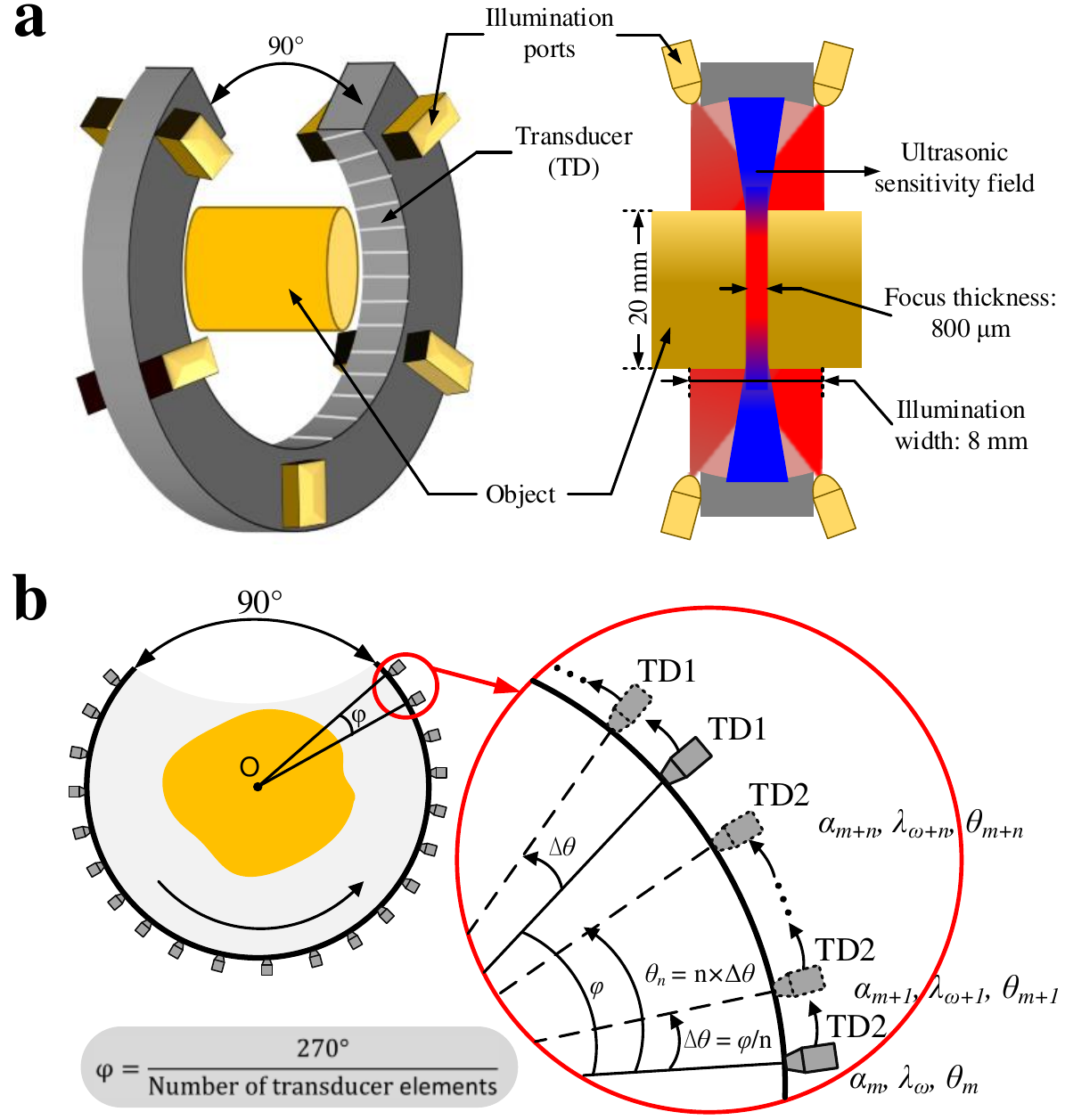}
\caption{(a) Schematic of the detection geometry of the commercial PAT system used in this study. (b) The detailed transducer layout of U3S-PAT simulated on the used PAT system. $\alpha$, $\lambda$, and $\theta$ indicate the scanning slice position, excitation wavelength, and transducer rotation angle, respectively.}
\end{figure}

\begin{figure*}[!h]
\centering
\includegraphics[scale=.33]{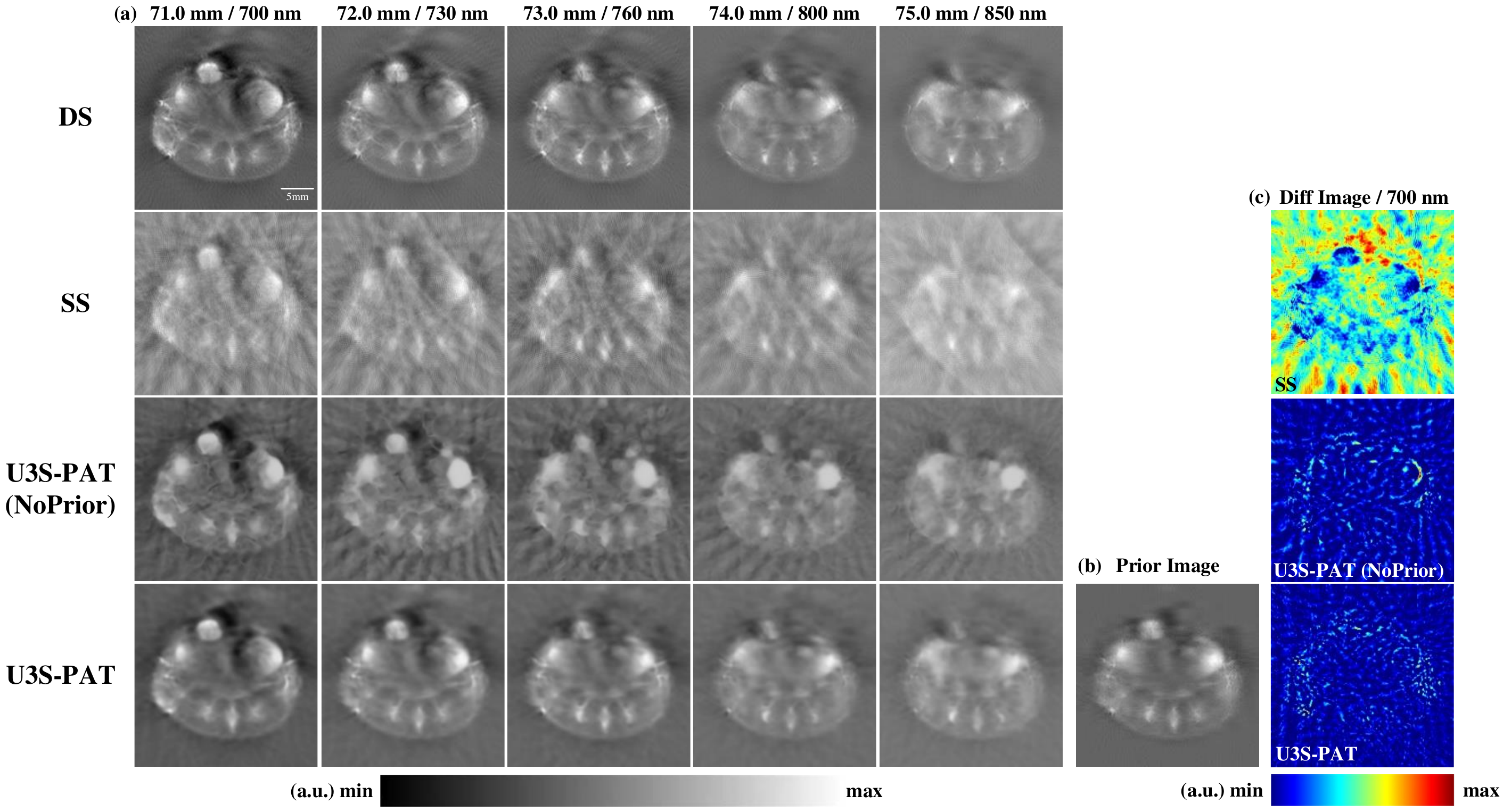}
\caption{The image reconstruction results of $\rm Nd = 21$ transducer: (a) PAT images obtained by different methods at different slice positions and wavelengths. (b) The structural prior image obtained by U3S-PAT. (c) Absolute normalized error map between the image reconstructed by each method and the reference dense sampled image.}
\end{figure*}

\begin{figure*}[!h]
\centering
\includegraphics[scale=.33]{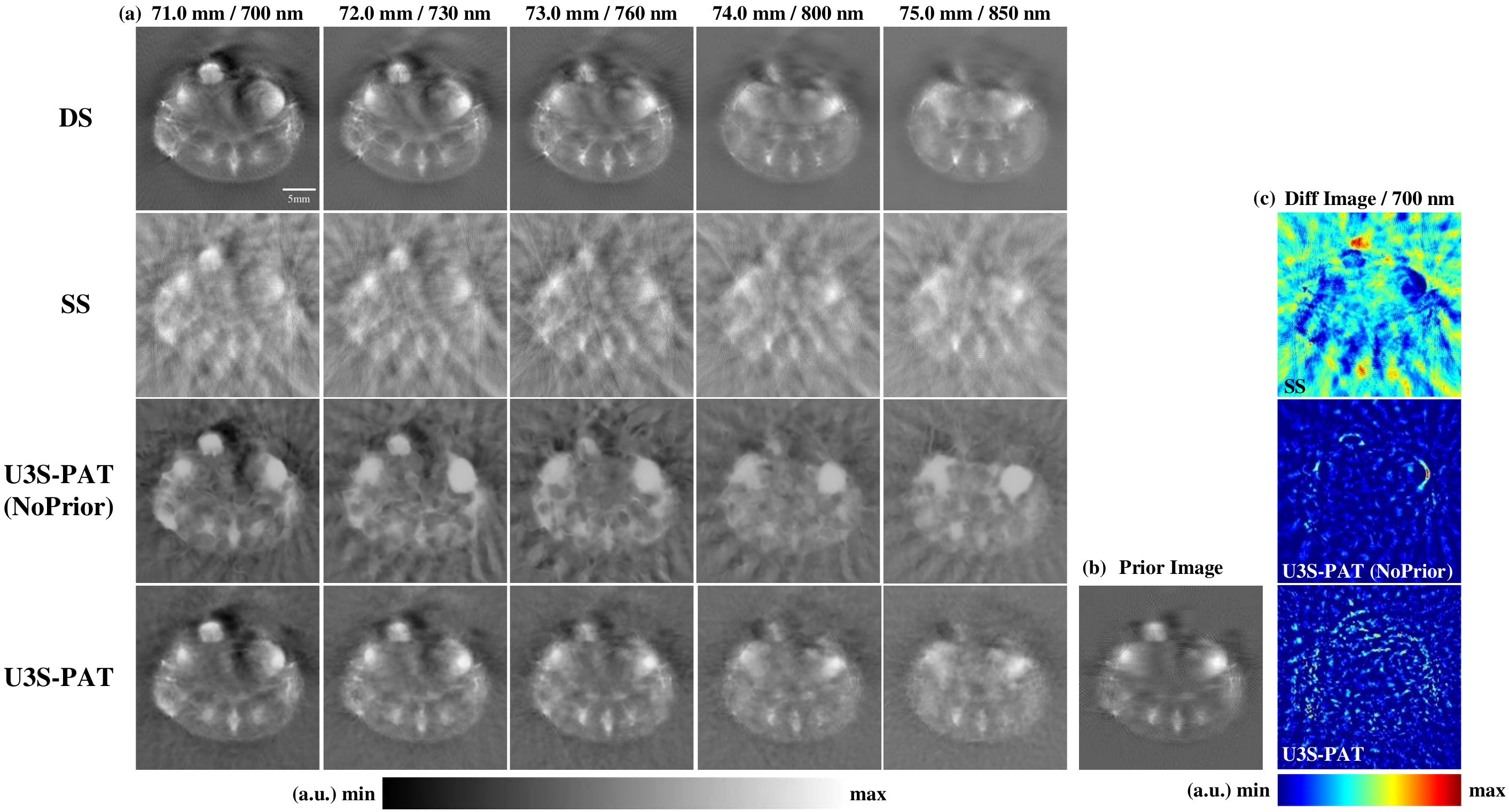}
\caption{The image reconstruction results of $\rm Nd = 16$ transducer: (a) PAT images obtained by different methods at different slice positions and wavelengths. (b) The structural prior image obtained by U3S-PAT. (c) Absolute normalized error map between the image reconstructed by each method and the reference dense sampled image.}
\end{figure*}

To showcase the effectiveness of the U3S-PAT method, the imaging configuration of the U3S-PAT scheme is simulated by leveraging a densely sampled dataset acquired by our MSOT system. By selectively retaining the signals acquired only at the sparse transducer locations and discarding the remaining densely sampled data, the spirally scanned and sparsely sampled data can be obtained. We use $\rm W = 5$ wavelengths for U3S-PAT in all our experiments, including 700 nm, 730 nm, 760 nm, 800 nm and 850 nm. We perform separate image reconstructions for two distinct numbers of transducer elements. As shown in Fig.3 (b), one transducer array consists of $\rm Nd = 21$ elements, corresponding to an element down sampling rate of $\sim$1/6, with a rotation angle of $\mathrm{\Delta}\theta = 2.57{^\circ}$ per step. The other transducer array consists of $\rm Nd = 16$ elements, corresponding to an element down sampling rate of 1/8, with a rotation angle of $\mathrm{\Delta}\theta = 3.37{^\circ}$ per step. With the above setup, the fused prior image used during image reconstruction covers the whole angular spacing.

In the U3S-PAT imaging system, the transducer array is horizontally translated and rotated during the scanning process. The translation distance in spectral unmixing experiments, also known as the slice spacing, is set to 0.5 mm, which is smaller than the 800 $\upmu$m focus detection sensitivity field of the laser. This scheme ensures that anatomical information is shared between neighboring images. While the translation distance in the image reconstruction experiments is set to 1.0 mm,  because the focal zone of each transducer element is characterized by the -6 dB zone (i.e. the area in which the sensitivity drops to half its maximum value), and the actual detection region is larger than the focus detection sensitivity region. 

\subsection{Small animal imaging experiments}
\textit{In vivo} animal experiments are conducted using healthy female nude mice (8 weeks old) on the commercial PAT system described in Fig.3. All animal experimental procedures are approved by the animal ethics committee of Southern Medical University and conducted following current guidelines. To minimize image artifacts caused by respiratory movements, the nude mice are anesthetized. Throughout the data acquisition process, the imaging chamber is filled with water and maintained at a constant temperature of 37°C. The speed of sound in the medium is set to 1536 m/s.

\hypersetup{hidelinks} 

\subsection{Implementation of image reconstruction}
In our experimental setup, all algorithms \footnote{https://github.com/zhongniuniu/U3S-PAT} are implemented using the PyTorch framework. The size of the reconstructed PAT image is defined as 220 × 220 pixels. For the prior embedding process, the training objective defined in Eq. (2) is optimized using the Adam optimizer, with a learning rate of 0.0001, over a total of 2000 training iterations. The reconstruction network is trained optimally, with the MLP network initialized using the prior embedding, and the objectives stated in Eq. (3) are trained for 2000 iterations using the Adam optimizer, with a learning rate of 0.00001. The hyperparameter $\delta$ in Eq. (5) is set to 0.8. 

\section{Experimental results}
\subsection{PAT images reconstruction at different locations under five wavelengths}
Fig.4 (a) and Fig.5 (a) display the image reconstruction results and the densely sampled (DS) reference images at different slice position and different wavelength. To investigate the impact of prior embedding, we also obtain the reconstruction results of U3S-PAT without the use of the prior image, i.e., random initialization is used during network training. As can be seen, both the sparsely sampled (SS) and U3S-PAT (NoPrior) methods suffer from significant background artifacts at the two transducer settings. In contrast, our U3S-PAT method effectively eliminates these artifacts, yielding reconstructed images of much better quality with well-defined anatomical structures and organ boundaries. When comparing Fig.4 (a) and Fig.5 (a), it could be seen that with even fewer transducer elements, the proposed method still produced acceptable results. In contrast, the image quality of the SS and U3S-PAT (NoPrior) reconstructions degraded heavily as the number of transducer elements decreased.

By comparing U3S-PAT with or without prior embedding, we observe that the introduction of prior significantly enhances image sharpness and reduces artifacts. Notably, despite utilizing the same prior image and relying solely on sparsely sampled data at each stage for image reconstruction, the resulting images accurately capture continuous tissue variations at different slice locations. To further observe the accuracy of the reconstructed images, absolute normalized error images between the reconstructed images obtained using each method and the corresponding reference images are calculated. Fig.4 (c) and Fig.5 (c) display the difference image between the reconstructed image at 21.0 mm / 700 nm and the corresponding DS reference. The U3S-PAT method exhibits the smallest error, emphasizing its superior performance compared to other methods. Moreover, comparing the results obtained from $\rm Nd = 21$ and $\rm Nd = 16$, the U3S-PAT images vary little with the number of transducer elements, while the absolute difference images of SS and U3S-PAT (NoPrior) became significantly worse as fewer transducer are used.

\begin{table*}[ht]
\renewcommand\arraystretch{1.4}
\caption{\label{tab1}PSNR $/$ SSIM OF IMAGE RECONSTUCTION RESULTS OBTAINED USING DIFFERENT METHODS. PSNR, PEAK SIGNAL TO NOISE RATIO; SSIM, STRUCTURAL SIMILARITY. (MEAN, \textbf{Bold}: BEST).}
\centering
\setlength{\tabcolsep}{5mm}{
\begin{tabular}{c|c|c|c|c|c}
\Xcline{1-6}{1.5pt}
\rule{0pt}{8pt} 
\textbf{Methods}           & \textbf{700 nm}         & \textbf{730 nm}         & \textbf{760 nm}         & \textbf{800 nm}         & \textbf{850 nm}         \\ \Xcline{1-6}{1.5pt} 
\multicolumn{6}{c}{\textbf{Number of transducer elements = 21}}                                                         \\ \hline \rule{0pt}{8pt}
SS                & 14.51 / 0.5814 & 16.72 / 0.6461 & 19.40 / 0.6565 & 17.91 / 0.7387 & 13.71 / 0.7318 \\ \hline \rule{0pt}{8pt}
U3S-PAT (NoPrior) & 30.60 / 0.7741 & 28.22 / 0.7068 & 28.71 / 0.7191 & 31.53 / 0.8099 & 33.50 / 0.8510 \\ \rule{0pt}{8pt}
\textbf{U3S-PAT}           & \textbf{34.76 / 0.8767} & \textbf{36.06 / 0.9011} & \textbf{35.45 / 0.8911} & \textbf{37.61 / 0.9225} & \textbf{37.47 / 0.9143} \\ \Xcline{1-6}{1.5pt}
\multicolumn{6}{c}{\textbf{Number of transducer elements = 16}}                                                         \\ \hline \rule{0pt}{8pt}
SS                & 13.67 / 0.5254 & 16.34 / 0.5962 & 17.57 / 0.5871 & 17.10 / 0.6532 & 17.15 / 0.6702 \\ \hline \rule{0pt}{8pt}
U3S-PAT (NoPrior) & 28.71 / 0.7196 & 26.50 / 0.6659 & 26.26 / 0.6522 & 28.83 / 0.7489 & 30.41 / 0.7824 \\ \rule{0pt}{8pt}
\textbf{U3S-PAT}           & \textbf{33.06 / 0.8362} & \textbf{33.87 / 0.8536} & \textbf{33.17 / 0.8356} & \textbf{34.18 / 0.8528} & \textbf{34.42 / 0.8442} \\
\Xcline{1-6}{1.5pt}
\end{tabular}}
\end{table*}

To further characterize the performance of U3S-PAT, we show the enlarged details of the images obtained using different methods in Fig.6. Upon observation of the magnified images, our U3S-PAT method exhibits superior structural visibility compared to both the SS method and U3S-PAT without prior embedding. Even at Nd=16, the quality of the U3S-PAT image remains unaffected, with only minor edge blurring observed in the region indicated by the red arrow. These findings highlight the effectiveness of the proposed self-supervised image reconstruction algorithm.

Next, we perform quantitative evaluation of the image reconstruction results. We use peak signal noise ratio (PSNR) and structural similarity (SSIM) for evaluation, and the obtained results are presented in Table \uppercase\expandafter{\romannumeral1}. At $\rm Nd = 21$, the U3S-PAT results showed an average increase of 120.48$\%$ in PSNR and 34.31$\%$ in SSIM compared to the SS results. At ND=16, the PSNR increased by 106.15$\%$ and SSIM increased by 39.25$\%$ on average. This means that the intensity values obtained by the U3S-PAT method are closer to the reference DS image in both settings. In addition, when $\rm Nd = 21$ and $\rm Nd = 16$, U3S-PAT improves PSNR by 18.87$\%$ and 19.89$\%$, and SSIM by 16.70$\%$ and 18.30$\%$, respectively, compared to the U3S-PAT (NoPrior) method. U3S-PAT method consistently achieves superior performance across all metrics when utilizing prior embedding.

\begin{figure}[ht]
\centering
\includegraphics[scale=.36]{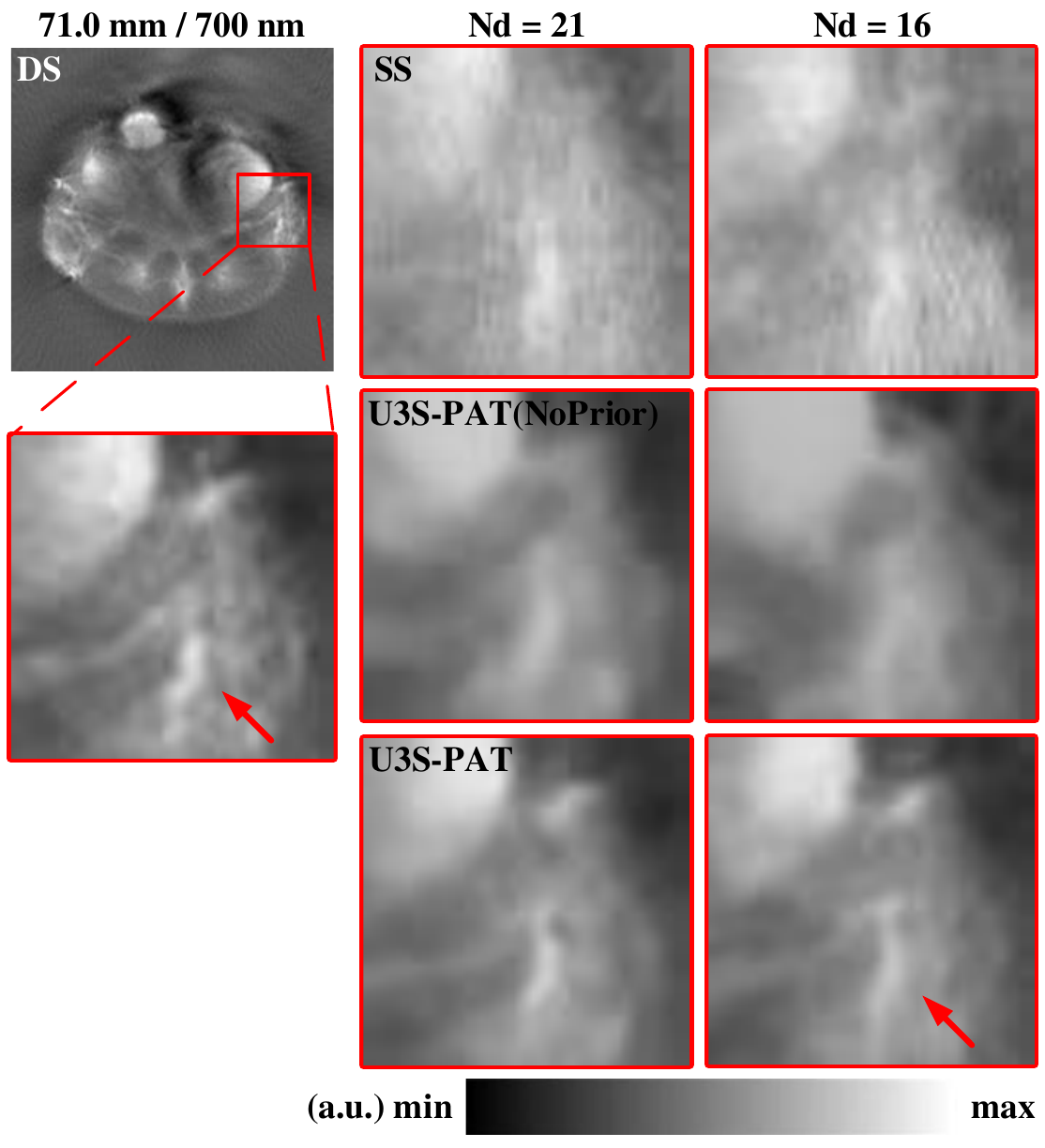}
\caption{Comparison of enlarged image reconstruction results of different transducer settings.}
\end{figure}

\begin{figure*}[!h]
\centering
\includegraphics[scale=.32]{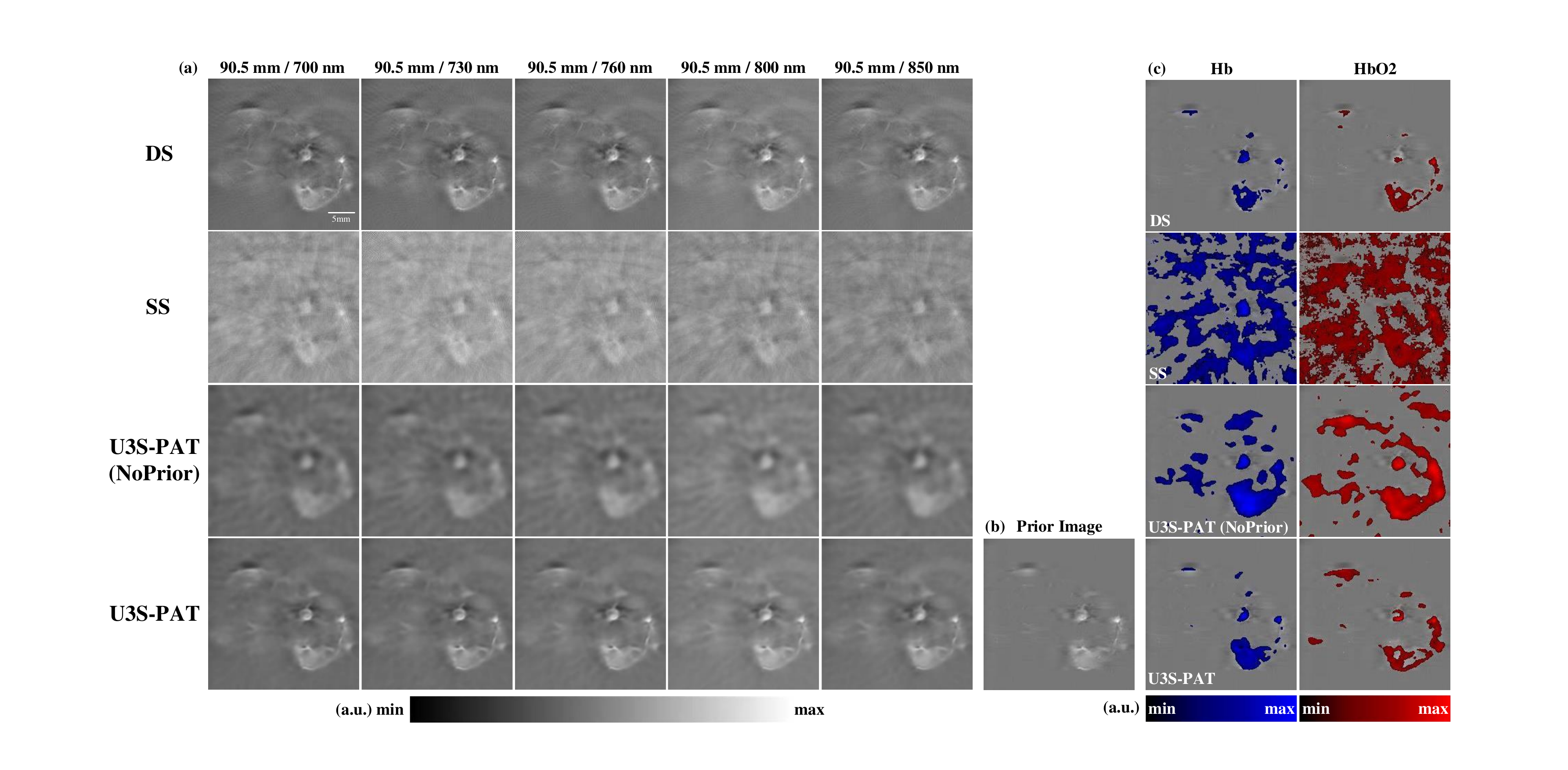}
\caption{The multispectral image reconstruction results at a single slice position (90.5 mm). (a) PAT images of five different wavelengths obtained by different methods. (b) The structural prior image obtained by U3S-PAT. (c) Spectral unmixing results of $\rm {HbO}_{2}$ and $\rm {Hb}$.}
\end{figure*}

\subsection{Spectral unmixing results}
Fig.7 shows the image reconstruction and spectral un-mixing results of the \textit{in vivo} animal experiment with $\rm Nd = 21$. The images reconstructed by different methods are depicted in Fig.7 (a). Analysis of the images reconstructed using the SS and U3S-PAT (NoPrior) methods reveals the presence of numerous artifacts, attributable to inadequate measurement angle and the absence of prior constraints. In contrast, the images obtained with the U3S-PAT method exhibit cleaner results, closely resembling the DS image. Fig.7 (c) illustrates the distribution images of $\rm {HbO}_{2}$ and $\rm {Hb}$, obtained by applying the linear spectral unmixing algorithm to each set of images. The spectral unmixing outcomes highlight the influence of artifacts in the reconstructed images on the unmixing accuracy. The results achieved by the U3S-PAT method align closely with the DS method, enabling clear differentiation of the distributions of $\rm {HbO}_{2}$ and $\rm {Hb}$ along organ contours.

The quantitative evaluation of the spectral unmixing experimental results is presented in Table \uppercase\expandafter{\romannumeral2}. The results demonstrate that the U3S-PAT method outperforms other methods in terms of PSNR, SSIM, and Dice metrics: compared to the SS method, the PSNR and SSIM increased for 100.50$\%$ and 27.00$\%$, respectively, and the Hb-Dice and HbO2-Dice increased for 202.39$\%$ and 320.59$\%$, respectively.

\begin{table}[ht]
\renewcommand\arraystretch{1.4}
\caption{\label{tab2}QUANTITATIVE ANALYSIS OF IMAGE RECONSTRUCTION AND SPECTRAL UNMIXING RESULTS OF DIFFERENT METHODS. Dice: DICE COEFFICIENT. (MEAN, \textbf{Bold}: BEST).}
\centering
\begin{tabular}{c|c|c|c|c}
\Xcline{1-5}{1.5pt}
\textbf{Methods}          & \textbf{PSNR}  & \textbf{SSIM}   & \textbf{Hb-Dice} & \textbf{HbO2-Dice} \\ \Xcline{1-5}{1.5pt}
SS               & 17.91 & 0.7180 & 0.2926  & 0.1986    \\ \hline
U3S-PAT (NoPrior) & 32.27 & 0.8482 & 0.7656  & 0.6530    \\
\textbf{U3S-PAT}          & \textbf{35.91} & \textbf{0.9119} & \textbf{0.8848}  & \textbf{0.8356}   \\ \Xcline{1-5}{1.5pt}
\end{tabular}
\end{table}

\subsection{Ablation studies}
\subsubsection{Impact of network structure on image reconstruction performance}
Selecting an appropriate network structure is crucial to ensure the optimal performance of the proposed U3S-PAT strategy. The function is influenced by the network's depth and width, which refer to the number of layers and neurons per layer, respectively. We obtained reconstruction results for the MLP network with various depths and widths at $\rm Nd = 21$ and slice spacing $= 1.0$ mm. The results are summarized in Table \uppercase\expandafter{\romannumeral3}. From the table, it can be observed that increasing the number of layers may cause inadequate optimization and lead to inferior reconstruction outcomes, and a certain increase in the width of the network can improve the performance of our reconstruction. Therefore, we selected the structure of a 4-layer MLP with 512 neural nodes as the backbone for our network.

\begin{table}[ht]
\renewcommand\arraystretch{1.4}
\caption{\label{tab3}QUANTITATIVE ANALYSIS OF IMAGE RECONSTRUCTION RESULTS UNDER DIFFERENT NETWORK STRUCTURES. (MEAN, \textbf{Bold}: BEST).}
\centering
\begin{threeparttable} 
\setlength{\tabcolsep}{5mm}{
\begin{tabular}{c|c|c}
\Xcline{1-3}{1.5pt}
\multicolumn{2}{c|}{\textbf{Network   Structure}} & \textbf{PSNR / SSIM ($\rm Nd=21$)} \\ \Xcline{1-3}{1.5pt}
\multirow{3}{*}{Width =512} & 4 Layers  & \textbf{36.65 / 0.9132}          \\ \cline{2-3}
                            & 6 Layers  & 33.12 / 0.7956          \\ \cline{2-3}
                            & 8 Layers  & 36.01 / 0.8989          \\ \Xcline{1-3}{1.5pt}
\multirow{3}{*}{Width =256} & 8 Layers  & 36.27 / 0.9011          \\ \cline{2-3}
                            & 16 Layers & 26.83 / 0.4765          \\ \cline{2-3}
                            & 20 Layers & 22.98 / 0.2715         \\ \Xcline{1-3}{1.5pt}
\end{tabular}}
\end{threeparttable} 
\end{table}

\subsubsection{Impact of slice spacing on image reconstruction performance} 
Selecting an appropriate slice spacing is essential to balance the data volume and quality of reconstructed images. To evaluate its impact on image quality, we reconstructed PAT images at different slice spacing and the results are depicted in Fig.8. As can be seen, as the slice spacing increases, the reconstructed images become increasingly blurred and artifacts emerge. The blurring can also be seen in the prior images, which confirms that the feasibility of our U3S-PAT strategy comes from redundant spatial information shared between adjacent slices during spiral scanning.

\begin{figure}[ht]
\centering
\includegraphics[scale=.31]{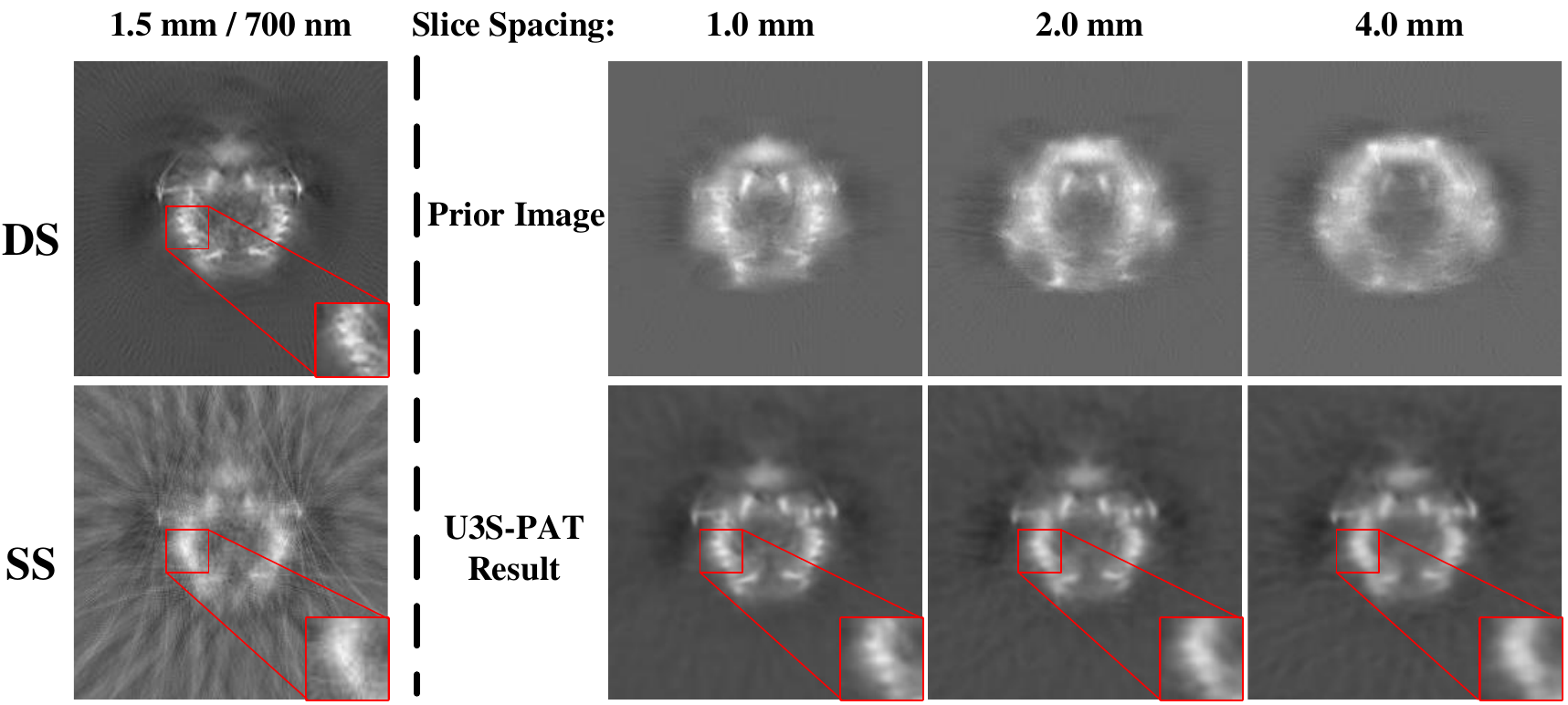}
\caption{Comparison of image reconstruction results of different slice spacings.}
\end{figure}

\subsubsection{Impact of number of transducer elements on image reconstruction performance}
To analyze the impact of element down sampling rate on the reconstruction performance, we conducted PAT image reconstructions with varying numbers of transducer elements. We have tested transducer settings of Nd = 4, 8, 16, 21, 26, 32, and 64 elements, and the obtained quantitative PSNR and SSIM results with respect to the transducer with Nd=128 are presented in Fig.9. As can be seen, as the number of transducer elements used in the measurement increases, both PSNR and SSIM increase accordingly. Specifically, optimal image quality is achieved when Nd is larger than 21. When a smaller number of elements is used, both PSNR and SSIM decrease rapidly.

\section{Discussion}
One of the primary drawbacks of multispectral PAT lies in the dense signal acquisition that necessitates high performance detector and system. The presented U3S-PAT method offers a new approach to address the challenge by proposing a novel spiral sparse sampling strategy, thus enables multispectral 3D signal acquisition with far less measurement compared to traditional dense sampling PAT. It is able to alleviate the need for multi-channel parallel signal acquisition and may significantly reduce system cost. Moreover, although has not been confirmed experimentally in our current study, our method is able to reduce the multispectral imaging time dramatically.

\begin{figure}[ht]
\centering
\includegraphics[scale=.21]{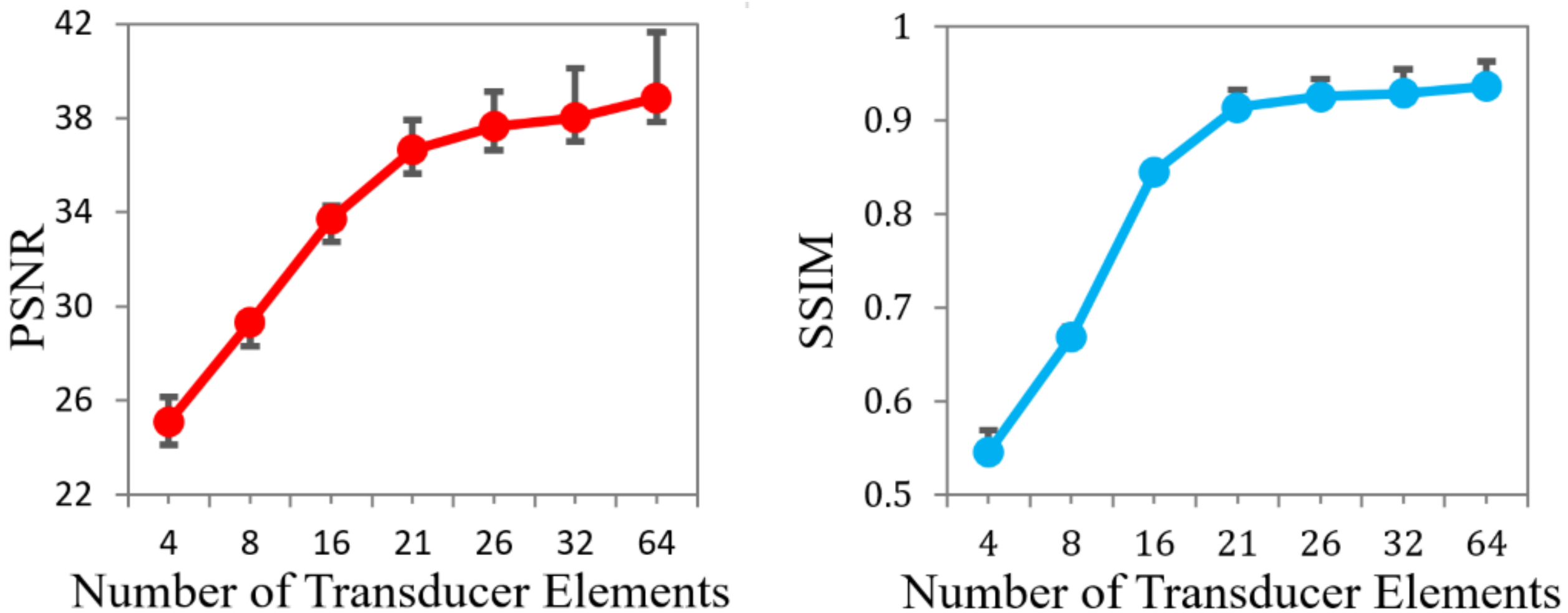}
\caption{PSNR and SSIM of the image reconstruction results obtained at different number of transducer elements.}
\end{figure}

The success of our method is based on the full use of spatial-spectral redundancy information during spiral scanning, which is induced to the image reconstruction of U3S-PAT. Our self-supervised neural network model eliminates the need for ground truth datasets and only relies on the raw acquired signal. The inclusion of the unique structural prior information provided by our U3S-PAT strategy significantly improves the reconstruction outcomes. Notably, our method can be used to further improve the imaging capability of dense sampling systems because the same wavelength-interlaced spiral scanning principle can also be applied on these systems.

The performance of the proposed U3S-PAT method is confirmed on a series of \textit{in vivo} animal experiments. However, in our current study, our U3S-PAT idea is validated based on signal simulated on a commercial transducer-fixed multispectral PAT system, its feasibility needs to be evaluated on a hardware-ready system in the future. The other limitation of our U3S-PAT method is that it is based on cross-sectional PAT geometry with a ring-shaped transducer. Its applicability on linear or spherical transducer settings has to be investigated.

\section{Conclusion}
In this work, we propose a novel ultra-sparse spiral sampling strategy for multispectral PAT. The strategy, which we named U3S-PAT, incorporates a novel spiral scanning principle with wavelength and transducer angle multiplexed based on a sparse ring-shaped transducer. We develop a self-supervised image reconstruction model that incorporates structural prior knowledge of spiral scanning to enhance image quality. The effectiveness of our U3S-PAT strategy is demonstrated through \textit{in vivo} animal experiments acquired on a commercial PAT system, and a total sparse sampling rate down to 1/30 is demonstrated. Our U3S-PAT provides a promising solution for the development of more efficient 3D multispectral PAT technology.

\appendices


\begin{thebibliography}{00}

\bibitem{b1}L. V. Wang and S. Hu, “Photoacoustic Tomography: \textit{in vivo} Imaging from Organelles to Organs,” \emph{Science}, vol. 335, no. 6075, pp. 1458–1462, Mar. 2012.

\bibitem{b2}L. V. Wang and J. Yao, “A practical guide to photoacoustic tomography in the life sciences,” \emph{Nat. Methods}, vol. 13, no. 8, pp. 627–638, Jul. 2016.

\bibitem{b3}P. K. Upputuri and M. Pramanik, “Recent advances toward preclinical and clinical translation of photoacoustic tomography: a review,” \emph{J. Biomed. Opt.}, vol. 22, no. 4, p. 041006, Nov. 2016.

\bibitem{b4}P. K. Upputuri and M. Pramanik, “Dynamic \textit{in vivo} imaging of small animal brain using pulsed laser diode-based photoacoustic tomography system,” \emph{J. Biomed. Opt.}, vol. 22, no. 09, p. 1, Sep. 2017.

\bibitem{b5}K. Tang et al., “Advanced Image Post-Processing Methods for Photoacoustic Tomography: A Review,” \emph{Photonics}, vol. 10, no. 7, p. 707, Jul. 2023.

\bibitem{b6}Y. Zhong et al., “Unsupervised Fusion of Misaligned PAT and MRI Images via Mutually Reinforcing Cross-Modality Image Generation and Registration,” \emph{IEEE Trans. Med. Imaging}, pp. 1–1, Jan. 2024.

\bibitem{b7}D. Razansky et al., “Multispectral opto-acoustic tomography of deep-seated fluorescent proteins \textit{in vivo},” \emph{Nat. Photonics}, vol. 3, no. 7, pp. 412–417, Jul. 2009.

\bibitem{b8}X. Luís Deán-Ben, N. C. Deliolanis, V. Ntziachristos, and D. Razansky, “Fast unmixing of multispectral optoacoustic data with vertex component analysis,” \emph{Opt. Lasers Eng.}, vol. 58, pp. 119–125, Jul. 2014.

\bibitem{b9}J. Glatz, N. C. Deliolanis, A. Buehler, D. Razansky, and V. Ntziachristos, “Blind source unmixing in multi-spectral optoacoustic tomography,” \emph{Opt. Express}, vol. 19, no. 4, p. 3175, Feb. 2011.

\bibitem{b10}S. Tzoumas, N. Deliolanis, S. Morscher, and V. Ntziachristos, “Unmixing Molecular Agents From Absorbing Tissue in Multispectral Optoacoustic Tomography,” \emph{IEEE Trans. Med. Imaging}, vol. 33, no. 1, pp. 48–60, Jan. 2014.

\bibitem{b11}K. Tang et al., “Learning Spatially Variant Degradation for Unsupervised Blind Photoacoustic Tomography Image Restoration,” \emph{Photoacoustics}, vol. 32, pp. 100536–100536, Aug. 2023.

\bibitem{b12}S. Zhang et al., “MRI Information-Based Correction and Restoration of Photoacoustic Tomography,” \emph{IEEE Trans. Med. Imaging}, vol. 41, no. 9, pp. 2543–2555, Sep. 2022.

\bibitem{b13}L. Qi et al., “Photoacoustic Tomography Image Restoration With Measured Spatially Variant Point Spread Functions,” \emph{IEEE Trans. Med. Imaging}, vol. 40, no. 9, pp. 2318–2328, May 2021.

\bibitem{b14}Z. Liang et al., “Automatic 3-D segmentation and volumetric light fluence correction for photoacoustic tomography based on optimal 3-D graph search,” \emph{Med. Image Anal.}, vol. 75, pp. 102275–102275, Jan. 2022.

\bibitem{b15}L. Yao and H. Jiang, “Photoacoustic image reconstruction from few-detector and limited-angle data,” \emph{Biomed. Opt. Express}, vol. 2, no. 9, p. 2649, Aug. 2011.

\bibitem{b16}J. Meng et al., “High-speed, sparse-sampling three-dimensional photoacoustic computed tomography \textit{in vivo} based on principal component analysis,” \emph{J. Biomed. Opt.}, vol. 21, no. 7, p. 076007, Jul. 2016.

\bibitem{b17}Y. Wang et al., “Enhancing sparse-view photoacoustic tomography with combined virtually parallel projecting and U3S-PATially adaptive filtering,” \emph{Biomed. Opt. Express}, vol. 9, no. 9, p. 4569, Aug. 2018. 

\bibitem{b18}H. Yu and G. Wang, “Compressed sensing based interior tomography,” \emph{Phys. Med. Biol.}, vol. 54, no. 13, pp. 4341–4341, Jun. 2009.

\bibitem{b19}G.-H. Chen, J. Tang, and S. Leng, “Prior image constrained compressed sensing (PICCS): A method to accurately reconstruct dynamic CT images from highly undersampled projection data sets,” \emph{Med. Phys.}, vol. 35, no. 2, pp. 660–663, Jan. 2008.

\bibitem{b20}E. Y. Sidky and X. Pan, “Image reconstruction in circular cone-beam computed tomography by constrained, total-variation minimization,” \emph{Phys. Med. Biol.}, vol. 53, no. 17, pp. 4777–4807, Aug. 2008.

\bibitem{b21}S. Ravishankar and Y. Bresler, “MR Image Reconstruction From Highly Undersampled k-Space Data by Dictionary Learning,” \emph{IEEE Trans. Med. Imaging}, vol. 30, no. 5, pp. 1028–1041, May 2011. 

\bibitem{b22}N. Davoudi, X. L. Deán-Ben, and D. Razansky, “Deep learning optoacoustic tomography with sparse data,” \emph{Nat. Mach. Intell.}, vol. 1, no. 10, pp. 453–460, Sep. 2019.

\bibitem{b23}A. Hauptmann et al., “Model-Based Learning for Accelerated, Limited-View 3-D Photoacoustic Tomography,” \emph{IEEE Trans. Med. Imaging}, vol. 37, no. 6, pp. 1382–1393, Mar. 2018.

\bibitem{b24}X. Li et al., “Multispectral Interlaced Sparse Sampling Photoacoustic Tomography,” \emph{IEEE Trans. Med. Imaging}, vol. 39, no. 11, pp. 3463–3474, Oct. 2020.

\bibitem{b25}N. R. Mollet et al., “High-Resolution Spiral Computed Tomography Coronary Angiography in Patients Referred for Diagnostic Conventional Coronary Angiography,” \emph{Circulation}, vol. 112, no. 15, pp. 2318–2323, Oct. 2005.

\bibitem{b26}C. B. Ahn, J. H. Kim, and Z. H. Cho, “High-speed spiral-scan echo planar NMR imaging-I,” IEEE transactions on medical imaging, vol. 5, no. 1, pp. 2–7, 1986.

\bibitem{b27}E. C. Ford, G. S. Mageras, E. Yorke, and C. C. Ling, “Respiration-correlated spiral CT: A method of measuring respiratory-induced anatomic motion for radiation treatment planning,” \emph{Med. Phys.}, vol. 30, no. 1, pp. 88–97, Dec. 2002.

\bibitem{b28}T. Fuchs, M. Kachelrieß, and W. A. Kalender, “Technical advances in multi–slice spiral CT,” \emph{Eur. J. Radiol.}, vol. 36, no. 2, pp. 69–73, Nov. 2000.

\bibitem{b29}X. L. Deán-Ben, T. F. Fehm, S. J. Ford, S. Gottschalk, and D. Razansky, “Spiral volumetric optoacoustic tomography visualizes multi-scale dynamics in mice,” \emph{Light Sci. Appl.}, vol. 6, no. 4, pp. e16247–e16247, Nov. 2016.

\bibitem{b30}A. Ron, X. L. Deán-Ben, J. Reber, V. Ntziachristos, and D. Razansky, “Characterization of Brown Adipose Tissue in a Diabetic Mouse Model with Spiral Volumetric Optoacoustic Tomography,” \emph{Mol. Imag. Biol.}, vol. 21, no. 4, pp. 620–625, Nov. 2018.

\bibitem{b31}S. K. Kalva, X. L. Deán-Ben, M. Reiss, and D. Razansky, “Head-to-tail imaging of mice with spiral volumetric optoacoustic tomography,” \emph{Photoacoustics}, vol. 30, p. 100480, Apr. 2023.

\bibitem{b32}B. Mildenhall, P. P. Srinivasan, M. Tancik, J. T. Barron, R. Ramamoorthi, and R. Ng, “NeRF: Representing Scenes as Neural Radiance Fields for View Synthesis,” in \emph{Proc. Europ. Conf. Comp. Visi(ECCV)}, pp. 405–421, 2020. 

\bibitem{b33}S. M. A. Eslami et al., “Neural scene representation and rendering,” \emph{Science}, vol. 360, no. 6394, pp. 1204–1210, Jun. 2018.

\bibitem{b34}M. Tancik et al., “Fourier Features Let Networks Learn High Frequency Functions in Low Dimensional Domains,” in \emph{Proc. Adv. Neural Inf.Process. Syst. (NeurIPS)}, Jun. 2020. 

\bibitem{b35}V. Sitzmann, J. N. P. Martel, A. W. Bergman, D. B. Lindell, and G. Wetzstein, “Implicit Neural Representations with Periodic Activation Functions,” in \emph{Proc. Adv. Neural Inf.Process. Syst. (NeurIPS)}, Jun. 2020.

\end{thebibliography}
\end{document}